\documentclass{article}

\usepackage{PRIMEarxiv}

\usepackage[utf8]{inputenc} 
\usepackage[T1]{fontenc}    
\usepackage{xcolor}
\usepackage{hyperref}       
\usepackage{url}            
\usepackage{booktabs}       
\usepackage{amsfonts}       
\usepackage{nicefrac}       
\usepackage{microtype}      
\usepackage{lipsum}
\usepackage{fancyhdr}       
\usepackage{graphicx}       
\graphicspath{{figure-and-tables/}}     

\title{Toward Template-Free Explainability for Monte Carlo Tree Search}

\author{
\begin{tabular}{ccc}
Siqi Lu$^{1}$ & Mirsaleh Bahavarnia$^{2}$ & Hiba Baroud$^{2}$ \\
Yixuan Zhang$^{1}$ & Hemant Purohit$^{3}$ & Ayan Mukhopadhyay$^{1*}$
\end{tabular}
\\[2em]
$^{1}$The College of William \& Mary \qquad
$^{2}$Vanderbilt University \qquad
$^{3}$George Mason University
}

\begin{document}
\maketitle

\begingroup
    \renewcommand{\thefootnote}{*}
    \footnotetext{Corresponding author: Ayan Mukhopadhyay (\href{mailto:amukhopadhyay@wm.edu}{\texttt{amukhopadhyay@wm.edu}})}
\endgroup

\begin{abstract}
    Probabilistic search algorithms, such as Monte Carlo Tree Search (MCTS), have proven very effective in solving sequential decision-making tasks under uncertainty. However, interpreting asymmetric search trees that incorporate bandit-based tree traversal and simulation-based value estimation is difficult for end users based solely on raw tree statistics. While prior work requires hand-crafted formal logic constraints that must be updated when the problem changes, we present a framework that enables large language models (LLMs) to generate evidence-grounded explanations of MCTS decisions from recorded search traces in an end-to-end manner. Our framework maps natural-language questions to a structured set of intent categories, determines whether the existing tree contains sufficient evidence, triggers targeted expansion when needed, and generates explanations using tree statistics such as visit counts, value estimates, and risk information. Experimental results 
    provide the first evidence that LLMs can serve as end-to-end explainers for probabilistic search, without requiring intermediate formal representations.
\end{abstract}

\section{Introduction}
    Many real-world problem settings include sequential decision-making under uncertainty formulated as stochastic control processes, e.g., transportation~\cite{sivagnanam2022offline}, emergency response~\cite{pettet2021hierarchical}, and healthcare~\cite{yu2021reinforcement}. A canonical way to mathematically represent such problems is a Markov decision process (MDPs)~\cite{puterman2014markov}. A popular class of algorithmic approaches to \textit{solve} an MDP is probabilistic search, which constructs a search tree, dictated by a model of the environment. Arguably, the most popular stochastic search method is Monte Carlo Tree Search (MCTS)~\cite{kocsis2006bandit}, which builds asymmetric search trees using a bandit-based approach that balances exploration of new trajectories and the exploitation of promising ones. As a result, MCTS has been applied in various domains, such as game playing~\cite{AlphaGo}, navigation~\cite{pitanov2023monte}, and planning~\cite{6145622,swiechowski2023monte}.

    The popularity of MCTS is driven by its key properties: it is \textit{anytime}, i.e., it returns a feasible action when interrupted, guaranteed to converge (given infinite time), and can naturally adapt to changing conditions given an updated model~\cite{bagchi2026digital,luo2024act}. However, one major challenge that hampers its adoption is its lack of interpretability~\cite{Baier2021-ck}. This is somewhat counterintuitive, since tree-based approaches are naturally interpretable. However, given the asymmetric nature of the search tree, the use of bandit-based heuristics, and the reliance on a large number of simulations (usually via a random policy) to compute value estimates, MCTS trees are difficult to parse. In practice, users may only see the final selected action or low-level statistics computed from the tree, such as visit counts of the nodes in the tree (that represent states) and value estimates. These outputs do not directly explain why one action was chosen, why alternatives were rejected, or what might happen under a counterfactual scenario or decision. This lack of interpretability creates challenges 
    to design trustworthy Artificial Intelligence (AI) systems when considering 
    explainable AI and human-AI interaction 
    for facilitating decision support,  
    especially in stochastic settings where even a reasonable action may still lead to failure~\cite{Baier2021-ck,wells2021explainable}.

    Prior work on explainable probabilistic search has argued that effective explanations of MCTS require a systematic mapping between queries made by end-users, i.e., in natural language, and the underlying search process~\cite{Baier2021-ck}. Existing methods, such as LogiEx~\cite{LogiEx,CTL_explainer_ziyan}, have argued that formal logic can provide this bridge. For example, in LogiEx~\cite{LogiEx}, user questions are first classified into query types and then mapped to a hand-crafted formal logic interface, which is tailored according to the specific MDP instance and the user-defined metrics for explainability. The mapped representation is then evaluated against evidence from the MCTS search tree to generate a natural-language explanation by using a large language model (LLM). However, this grounding comes with a tradeoff: the system relies on predefined query categories, logic templates, domain-specific variables, and scorer functions to map user questions to search-tree evidence. As a result, adapting the framework to a new task or state/action space, even arising out of a small change to the MDP instance, requires additional expert effort to redesign these mappings~\cite{CTL_explainer_ziyan,LogiEx}.
    

    We argue that recent progress in LLMs suggests a different possibility~\cite{klissarov2024modeling,baier2026explanations}. Instead of requiring a fixed formal representation for every type of user question based on the MDP, we explore whether LLMs can answer evidence-grounded user queries to MCTS traces in an end-to-end manner through effective prompting, without requiring hand-designed domain-specific logic templates. In our proposed method, the MCTS tree remains the main source of information, but an LLM is used to interpret the users' questions, identify the relevant node, path, or general search behavior, and generate an explanation report grounded in the recorded search trace. If the existing tree does not contain enough evidence to answer the question (since users can ask questions about paths that the tree did consider promising), the framework can detect this gap and trigger targeted expansion. This allows the system to support flexible natural-language interaction while keeping explanations grounded in the actual MCTS process. 
    Our core contributions are:
    
    \begin{enumerate}
        \item We frame MCTS interpretability as a user-centered explainability problem, where users can ask natural-language questions about the reasoning that resulted in a decision computed by the search tree, instead of relying on final actions or low-level tree statistics.
        \item We propose an interactive LLM-based end-to-end framework that connects user questions with recorded MCTS traces and generates explanations grounded in the search process. \textit{To the best of our knowledge, this is the first framework that enables the explainability of probabilistic search trees without any intermediate formal representation.}
        \item We show how the framework can support different types of user questions, including why, why-not, and what-if questions, providing a foundation for future evaluation in more realistic user-facing settings.
    \end{enumerate}
\section{Related Work}
    \noindent \textbf{Explainable Sequential Decision-Making.} Rather than only presenting the final selected action, explanation methods aim to help users understand why an action was chosen, why alternatives were not selected, and what future outcomes may result from different decisions~\cite{wells2021explainable,samek2019explainable,arrieta2020explainable}. This is especially important in stochastic environments, where decision-making and outcomes can be non-intuitive to end-users~\cite{miller2019explanation}. Based on prior work, we focus on explanation methods that are grounded in the system’s actual decision process rather than only providing post-hoc summaries~\cite{sreedharan2020bridging,langley2016explainable}. For a deeper dive into this area, we refer readers to a recent survey on explainability for sequential decision-making~\cite{baier2026explanations}.\\
        
    \noindent \textbf{MCTS Interpretability}
        MCTS is difficult to interpret because its decisions are based on sampled future trajectories, visit counts, value estimates, and branching alternatives \cite{Baier2021-ck,baier2026explanations,wells2021explainable}. 
        More recent work, such as LogiEx \cite{LogiEx} and CTL(computation tree logic)-based explainers~\cite{CTL_explainer_ziyan}, combine LLMs with formal logic, domain knowledge, and structured evidence extraction to generate natural-language explanations for MCTS-based planning. While this approach provides strong grounding, it also relies on predefined query categories, logic templates, domain-specific variables, and scorer functions~\cite{LogiEx}. In contrast, our work investigates whether LLMs, by themselves, can provide a lightweight connection between natural-language user questions and recorded MCTS traces while still grounding explanations in the actual search process.

\begin{figure}
    \centering
    \includegraphics[width=\linewidth]{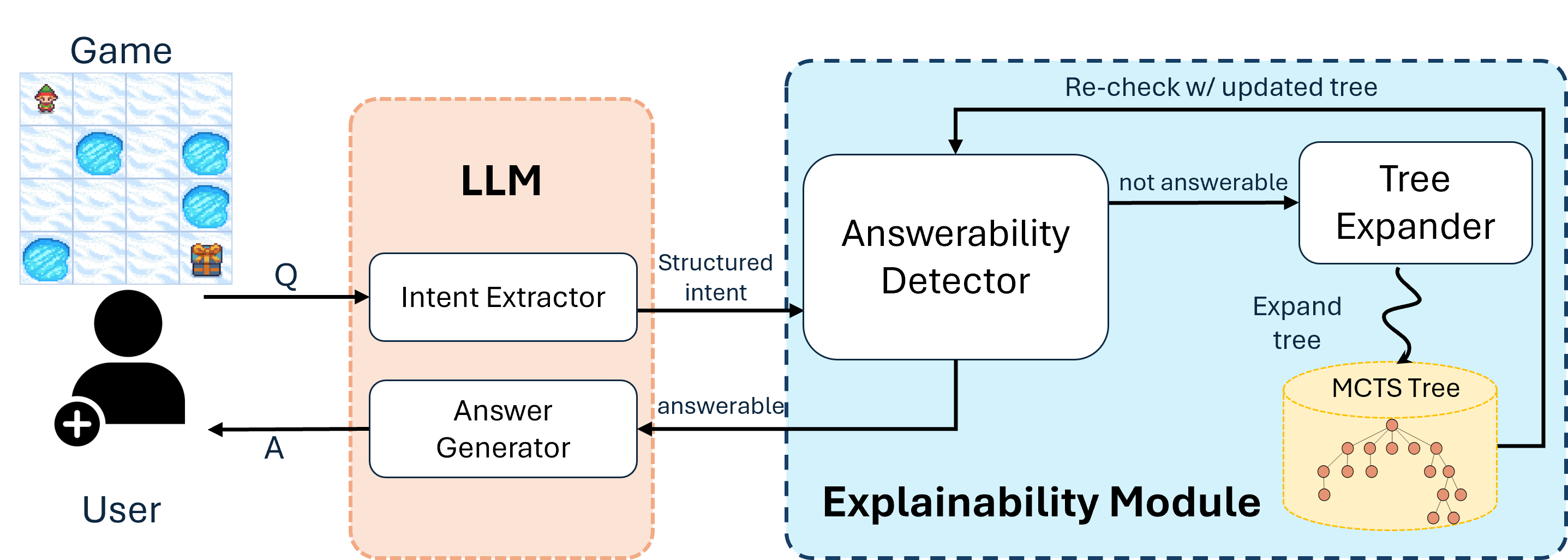}
    \caption{Overview of the proposed framework.}
    \label{fig:framework}
\end{figure}
\section{Methods}
We propose an interactive explanation framework for MCTS-based decision-making tasks. As shown in Figure~\ref{fig:framework}, the framework provides evidence-based explanations by linking user questions to the search tree generated during planning. Our framework assumes three \textit{actors}: 1) an end-user or decision-maker who has domain expertise in the decision-making problem but most likely does not have an understanding of probabilistic search (e.g., a transportation operator); 2) a computational agent or planner that invokes the search algorithm; and 3) the explainability module that we propose. Our goal is to enable (1) the end-user to use, (2) the computational agent or planner, and (3) the explainability module to understand.

Our explainability module takes two inputs: a question from the end-user in natural language and a saved MCTS tree that records visited states (denoted by nodes in the tree), available actions at each state, visit counts, and value estimates generated by rollouts (or a trained value estimator, e.g., a neural network). 
The LLM interprets the user's question, identifies the relevant part of the tree to base the explanation on (i.e., the node and the branch), checks whether the existing tree contains sufficient evidence, and generates a natural-language explanation grounded in the recorded search process. We provide guidance to the LLM through a detailed prompt. When the relevant state, action, or path is missing or insufficiently explored, the framework can perform targeted expansion from the corresponding node. The explainability module contains five main components: tree recording, intent extraction, answerability detection, targeted expansion, and answer generation.

\noindent \textbf{1) Recorded MCTS Tree}
    During planning, the computational agent runs MCTS to select an action in the environment. At each decision step, we save the search tree generated by MCTS. Each node in the tree represents an environment state, and each edge represents an action or stochastic transition. For each node, we record relevant search statistics, including visit counts, estimated values, available actions, parent-child relationships, and transition outcomes.\\

\noindent \textbf{2) User Query and Intent Extraction}
    Users may ask different types of questions about the agent’s decision-making process. Some questions focus on why the agent selected a specific action at a given state, while other questions can be counterfactual, asking what would have happened if the agent had chosen a different action. We do not put any restrictions on the questions that the user may ask.
    To connect these natural-language questions with the MCTS tree, we first convert the user query into a a structured representation of the intent category using a large language model (llama-3.3 \cite{grattafiori2024llama}). The structured representation of intent includes the question type, target state, target action, and, when applicable, the target path. This structured representation allows the following modules to locate the relevant part of the tree and determine what evidence is needed to answer the question.\\

\noindent \textbf{3) Answerability Detection}
    After extracting the structured intent, an answerability detector module checks whether the saved MCTS tree contains sufficient evidence to answer the user’s question, since the question might be about a part of the tree that the bandit-based expansion strategy deemed unpromising. In our current preliminary implementation, it is a human-in-the-loop business logic.
    For counterfactual or underexplored-action questions, if the existing tree does not have enough information, the detector identifies the missing or insufficient branch and marks the query as requiring additional expansion.\\

\noindent \textbf{4) Evidence Tree Expansion}
    When the evidence detector determines that the existing tree is insufficient, the framework performs targeted tree expansion. Instead of rerunning MCTS from the initial state, the expander invokes the computational agent (component (2) described above) to perform additional simulations, thereby preventing the system from generating explanations based on incomplete evidence. The tree is not generated from scratch to reduce latency; instead, the structured intent allows the framework to grow the tree from the specific node that the user's question is based on.\\
    
\noindent \textbf{5) Explanation Generation}
    Finally, the LLM generates a natural-language explanation using the collected tree evidence. The explanation module receives the original user question, the extracted intent, and the relevant MCTS statistics, such as visit counts, estimated values, selected actions, alternative actions, and transition risks.

\section{Evaluation}
\begin{figure}
    \centering
    \includegraphics[width=0.35\linewidth]{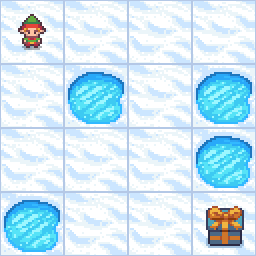}
    \caption{The FrozenLake environment used for evaluation. The environment presents the canonical challenge of sequential decision-making under uncertainty: the agent must navigate a slippery surface while avoiding the holes to claim the reward. Image Source: OpenAI’s Gym Library.}
    \label{fig:frozen-lake}
\end{figure}

While we acknowledge the need for a detailed human study for gauging the efficacy of our framework, we conducted a technical analysis for this preliminary evaluation. We evaluate whether each module can correctly extract intent, identify relevant search-tree evidence, detect answerability, and generate grounded explanations. We evaluate the framework in a stochastic FrozenLake \cite{towersgymnasium} environment (Figure~\ref{fig:frozen-lake}). In this canonical grid-world environment, the agent must navigate from a start state to a goal state while avoiding holes. Uncertainty arises from the frozen surface---the agent cannot always navigate in its intended direction of movement. At each decision step, MCTS selects an action and records the resulting search tree for later explanation. We use the LLM \textit{Llama-3.3-70b-versatile} \cite{grattafiori2024llama} in this section.

\noindent \textbf{Query set creation}
        We construct a handcrafted query set for the proof-of-concept evaluation. The query set contains 21 annotated question-tree pairs, where each sample includes a natural-language user question and a corresponding recorded MCTS tree file. For example, our questions include contrastive queries (\textit{``Why did the agent choose Up at the current state?''}), counterfactuals (\textit{``What would the agent's strategy look like if the Left action had been explored at the current state?''}), and questions about multiple actions taken over time (\textit{``Why does going Right then Down from state 13 lead most reliably toward the goal?''}). The questions cover three levels of explanation: node-level, path-level, and general questions about the agent’s search behavior. 
        For answerable questions, the required evidence is already available in the recorded tree. For non-answerable questions, the dataset specifies the state-action pair where targeted expansion should begin. These annotations allow us to evaluate intent extraction, answerability detection, expansion target selection, and explanation faithfulness separately.
        
\noindent \textbf{Component-Level Evaluation}        
        We evaluate each component of the overall framework. We focus on the intent extractor and the answerability detector; note that the targeted expander is not evaluated separately since the extracted intent and the detector output directly specify the expansion target.

        For the intent extractor, we compare the extracted structured intent with the manually annotated ground truth. The compared fields include the predicted question type, target state, target action, and target path when applicable. 
        At the field level, the extractor achieved 85.7\% accuracy for question type, 95.2\% for target state, 71.4\% for target action, and 90.5\% for target path (see Table~\ref{tab:intent-extraction}). These results suggest that the extractor can reliably identify states and paths, but action extraction remains more challenging. We are currently exploring providing more structured prompts to improve action extraction.

        \begin{table}[t]
\centering
\caption{Intent extraction results on the annotated query set.}
\label{tab:intent-extraction}
\begin{tabular}{lcc}
\hline
\textbf{Metric} & \textbf{Correct / Total} & \textbf{Accuracy} \\
\hline
Question type & 18 / 21 & 85.7\% \\
Target state & 20 / 21 & 95.2\% \\
Target action & 15 / 21 & 71.4\% \\
Target path & 19 / 21 & 90.5\% \\
\hline
\end{tabular}
\end{table}

        For the answerability detector, we evaluate whether the framework correctly determines if the existing MCTS tree contains sufficient evidence to answer the user query (shown in Table~\ref{tab:answerability-detection}). The detector correctly classified all 21 queries, achieving an accuracy of 100\%. 
        In future work, we plan to investigate whether this rule-based component can be replaced or complemented by an LLM-based detector. 

        \begin{table}[t]
\centering
\caption{Answerability detection results on the annotated query set.}
\label{tab:answerability-detection}
\begin{tabular}{lcc}
\hline
\textbf{Metric} & \textbf{Correct / Total} & \textbf{Accuracy} \\
\hline
Answerability decision & 21 / 21 & 100.0\% \\
\hline
\end{tabular}
\end{table}

\noindent \textbf{Explanation Faithfulness}
        In addition to the component-level checks, we manually evaluate the generated explanations. Specifically, we examine whether the final LLM answer mentions the key evidence required by the question, including the agent’s selected action, the user-specified action when applicable, and risk-related evidence. This check is not intended to replace a full human evaluation of explanation quality. Instead, it provides an initial indication of whether the generated explanation includes the core information needed to justify the MCTS decision.

        As shown in Table~\ref{tab:explanation-grounding}, across 21 queries, the generated explanations correctly mentioned the agent action in 15 cases and risk-related evidence in 19 cases. For user-specified actions, 3 queries were not applicable, and the explanations correctly mentioned the user action in 16 out of the remaining 18 cases. 
        These results suggest that the explanation generator often includes important MCTS evidence, especially risk-related information, but it may still omit some action-specific details.
        Since this analysis only checks whether key evidence is mentioned, it should be interpreted as an initial grounding check rather than a complete assessment of explanation usefulness or user trust.
                \begin{table}[t]
\centering
\caption{Keyword-based grounding results for generated explanations.}
\label{tab:explanation-grounding}
\begin{tabular}{lcc}
\hline
\textbf{Keyword Check} & \textbf{Passed / Total} & \textbf{Rate} \\
\hline
Agent Core Decision & 16 / 21 & 76.2\% \\
Risk Calculation & 19 / 21 & 90.5\% \\
Asked State-Action Pair & 19 / 19 & 100.0\% \\
All checks passed & 15 / 21 & 71.42\% \\
\hline
\end{tabular}
\end{table}

\begin{figure}[t]
    \centering
    \includegraphics[width=0.8\linewidth]{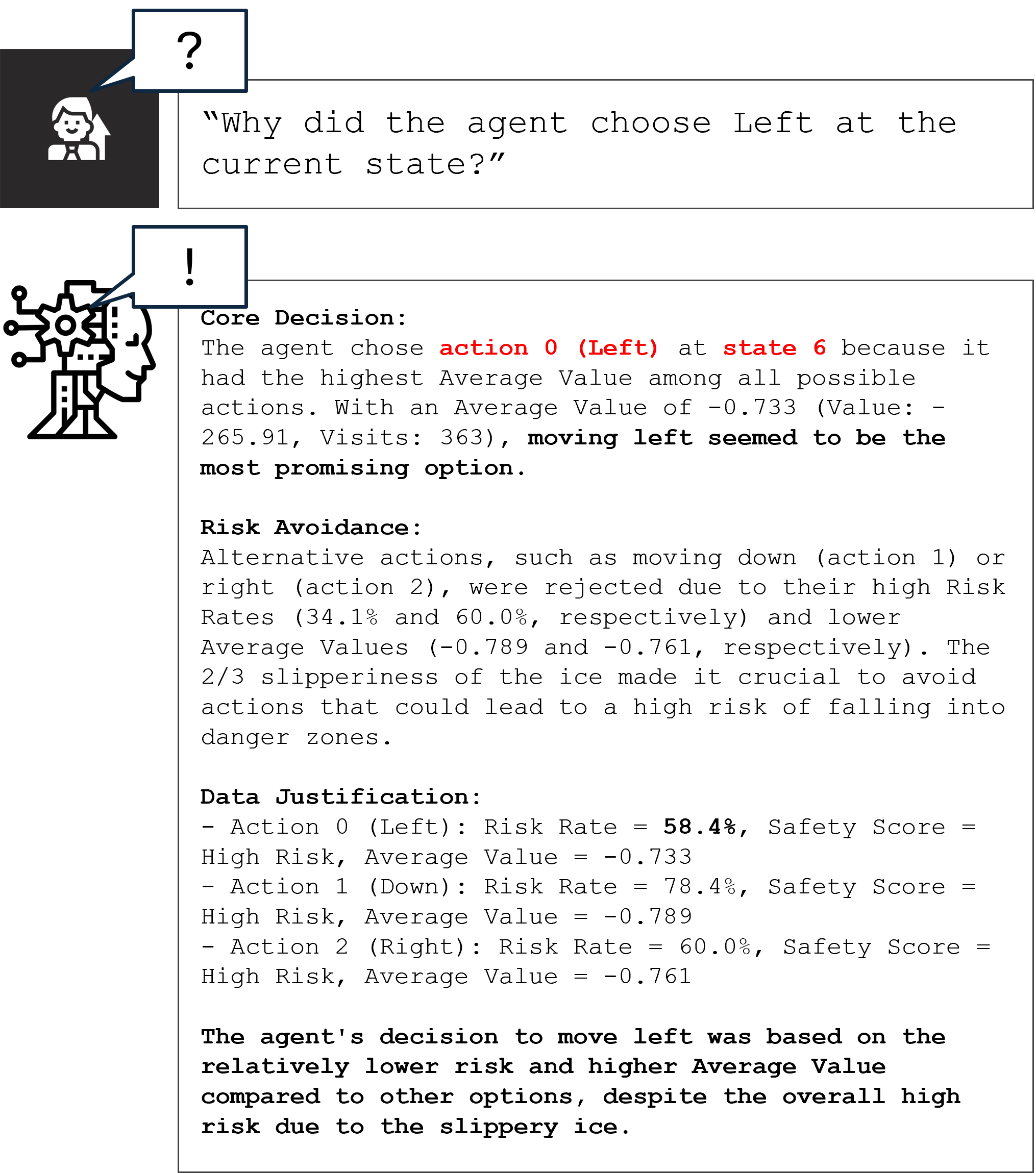}
    \caption{Qualitative example of the generated explanation.}
    \label{fig:example}
\end{figure}

\noindent \textbf{Qualitative Analysis}    
        To better understand the keyword-based grounding results, we present one representative explanation generated by our framework in Figure~\ref{fig:example}. The user asks "\textit{why did the agent chose Left at the current state}". The explanation grounds its answer in recorded MCTS evidence by identifying the selected action and state and reporting statistics such as value estimates, visit counts, and risk estimates.

        On the other hand, this case nevertheless exposes some limitations of the current generation step. First, the answer is relatively verbose and \textit{report-like}, which may reduce readability for users who ask a direct question. Second, although the current state contains multiple available actions, the explanation discusses only two alternatives and does not explain why the remaining actions are omitted. This creates an action coverage issue where the user cannot tell whether the missing action was unavailable, underexplored, or simply ignored in the explanation. The format and coverage of actions can be improved through more structured prompts. We suggest that these results should be viewed as a preliminary evidence. 

\section{Conclusion}
    In this paper, we present, to the best of our knowledge, the first end-to-end LLM-based explainer for probabilistic search. Our framework focuses on MCTS, arguably the most popular probabilistic search algorithm for sequential decision-making. We provide preliminary evidence that LLM-based frameworks can support different explanation scopes, including node-level, path-level, and general questions, without the need for intermediate formal representations, thereby increasing the flexibility and generalization of such frameworks. In our future work, we will explore the completeness and conciseness of generated explanations across different task domains, evaluate the targeted expansion module more deeply, and conduct user studies.
    

    


\bibliographystyle{unsrt}  
\bibliography{reference}  

\end{document}